\documentclass[letterpaper,aps,twocolumn,showpacs,superscriptaddress,
floats,prb]{revtex4}

\usepackage[english]{babel}
\usepackage[latin1]{inputenc}
\usepackage{amsmath, amssymb, amsfonts}
\usepackage{graphicx,epsfig,dsfont}
\usepackage{float}

\usepackage[squaren]{SIunits}
\usepackage{color}

\providecommand{\vect}[1]{\boldsymbol{#1}}

\begin{document}

 \title{Current induced rotational torques in the skyrmion lattice phase of 
chiral magnets}

\author{Karin~Everschor}
\author{Markus~Garst}
\affiliation{Institute of Theoretical Physics, University of Cologne, 
D-50937 Cologne, Germany
}
\author{R.~A.~Duine}
\affiliation{Institute for Theoretical Physics, Utrecht University, 3584 
CE Utrecht, The Netherlands}
\author{Achim~Rosch}
\affiliation{Institute of Theoretical Physics, University of Cologne, 
D-50937 Cologne, Germany
}

\begin{abstract}
In chiral magnets without inversion symmetry, the magnetic structure can form a lattice of magnetic whirl lines, a
two-dimensional skyrmion lattice, stabilized by spin-orbit interactions in a
small range of temperatures and magnetic fields. The twist of the magnetization
within this phase gives rise to an efficient coupling of macroscopic magnetic
domains to spin currents. We analyze the resulting spin-transfer effects, and,
in particular, focus on the current induced rotation of the magnetic texture by
an angle. Such a rotation can arise from macroscopic temperature gradients in
the system as has recently been shown experimentally and theoretically. Here we
investigate an alternative mechanism, where small distortions of the skyrmion
lattice and the transfer of angular momentum to the underlying atomic lattice
play the key role. We employ the Landau-Lifshitz-Gilbert equation and adapt the
Thiele method to derive an effective equation of motion for the rotational 
degree of freedom. 
We discuss the dependence of the rotation angle on the
orientation of the applied magnetic field and the distance to the phase
transition.
\end{abstract}

\date{\today}

\maketitle

\section{Introduction}

In a magnetic metal, angular momentum can be transferred from spins of the
conduction electrons to the magnetization and vice versa. 
In non-equilibrium conditions, this flow of angular momentum -- the so-called
spin-transfer torques  -- results in very interesting phenomena if the
magnetization is spatially non-uniform.\cite{Ralph08} For example, a
spin-polarized current is able to induce domain-wall motion in nano
wires,\cite{Beach:2005p8560,Hayashi:2007p8559} microwave oscillations in
magnetic multilayers,\cite{Kiselev:2003p8558,Boulle:2007p8557} or vortex
oscillations in magnetic nano pillars.\cite{Pribiag:2007p8556} 
The ability to control magnetic configurations by electric currents may have 
interesting applications for non-volatile magnetic memory.\cite{Parkin:2008p8497}

\begin{figure}[t]
\begin{center}
\includegraphics[width=8.5cm]{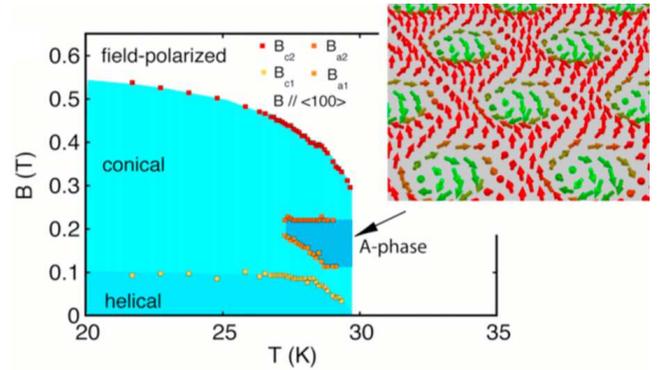}
\end{center}
\caption{Magnetic phase diagram of MnSi taken from Ref.~[\onlinecite{Jonietz10}]. At small magnetic fields close to the critical temperature $T_c =29.5$K a skyrmion lattice is stabilized, historically denoted as the A-phase. The inset shows the hexagonal arrangements of single skyrmions. 
\label{fig1}}
\end{figure}

Most experimental studies investigate spin-transfer torque effects in nano
structures, which will be important for the design of future memory devices.
In such nano structures, rather large current densities (typically larger than
$10^{11}$A/m$^2$) are needed to induce, e.g., the motion of domain walls 
but they can be applied in these systems without substantial Joule heating.
Recently, spin transfer torque effects at much smaller current densities 
($10^{6}$A/m$^2$)
have been observed by Jonietz {\it et al.} \cite{Jonietz10} with neutron
scattering in  a {\em bulk} sample of MnSi. In this material
a peculiar magnetic structure, a lattice of magnetic whirls or `skyrmions', is
stabilized in a small range of magnetic fields and temperatures, see
Fig.~\ref{fig1}. 
A rotation of this skyrmion lattice by a finite angle is observed experimentally
if the current density exceeds a critical threshold value.

MnSi is an example of a chiral magnetic metal. While its Bravais lattice is
cubic, the atomic structure (P2$_1$3) has no inversion symmetry. It is therefore
`chiral', i.e. the atomic crystal and its mirror image do not match. 
The chirality of the crystal implies that the magnetization 
likes to twist in this material due to relativistic effects (the Dzyaloshinskii-Moriya interaction), typically by forming a spiral. 
Close to the critical temperature and in the presence of a small magnetic field, a more complex magnetic structure is formed: a skyrmion lattice. Such a skyrmion lattice is in some aspects similar to the vortex lattice in a type-II superconductor. 
The magnetic structure is organized in a hexagonal lattice perpendicular to the magnetic field and is translationally invariant in parallel direction. While
in a superconductor the phase of the order parameter winds by $2 \pi$ around the core of each vortex
where the order parameter vanishes, the  winding of the magnetization is more complex, see Fig.~\ref{fig1}. The magnetization remains always finite but its direction winds once around a sphere
without any singular point. Such a configuration is topological stable and is called a `skyrmion' after the nuclear physicist Tony Skyrme, who showed in a pioneering work that certain configurations of pion fields have the same properties as baryons.\cite{Skyrme:1961p8505,Skyrme:1962p8506}

Experimentally, the hexagonal magnetic lattice in MnSi was detected by neutron
scattering.\cite{mueh09} The winding of the magnetization was identified by
an extra contribution to the Hall effect.\cite{neub09} Also other
materials with the same crystal symmetry show the same skyrmion phase,
e.g.  Fe$_{1-x}$Co$_{x}$Si.\cite{muen09,yu:2010,pfleiderer:JPCM2010}
In the latter material, the skyrmion structure has been directly
measured using Lorentz force microscopy by Yu {\it et al.}
\cite{yu:2010} Theoretically, it has been pointed out in a pioneering
early work of Bogdanov and Yablonskii\cite{bogd89} that Dzyaloshinskii
Moriya interactions in chiral magnets favor magnetic skyrmion
textures. Based on a mean-field analysis, it was, however,
argued\cite{bogd94} that in materials with the symmetry of MnSi
such a phase is never thermodynamically stable but only metastable. In
contrast, we showed theoretically in Ref.~[\onlinecite{mueh09}] that the
skyrmion phase becomes thermodynamically stable in a small temperature
window when thermal fluctuations are properly taken into
account. Interestingly, the skyrmion phase becomes stable even on the
mean field level in films with a small perpendicular magnetic
field.\cite{yu:2010} Skyrmion-like magnetic textures in chiral magnets
have, for example, also been considered in
Refs.~[\onlinecite{roes06,binz06,binz06b,binz08,fisc08,Yi09,Han10}].

From the perspective of spin-transfer torque, skyrmion lattice phases as in MnSi
are particular interesting
as the peculiar twist of the magnetization results in an efficient coupling of
currents and the magnetization. Whereas for traditional
spintronic devices such a coupling primarily occurs when the magnetization winds
in a domain wall or some nanoscopic device, it extends over the macroscopic bulk
phase for 
a skyrmion lattice. Indeed, the so-called gyrocoupling vector, introduced by 
Thiele \cite{Thiele72} many years ago to describe 
the motion of magnetic domains, becomes proportional to the volume as the gyrocoupling per volume can directly
be identified with the skyrmion density.
Physically, this coupling can either be visualized as arising from Berry phases
which an electron picks up when moving across a skyrmion texture
\cite{pfleiderer:Nature2010} or as a 
Magnus force arising from the interplay of external and circulating internal
spin currents.\cite{Jonietz10}

The above described gyrocoupling and further dissipative forces are expected to induce
a motion of the skyrmion lattice above a critical current strength determined by the pinning
of the magnetic structure by disorder. The resulting translational motion of the magnetic structure is, however, extremely difficult to observe with neutron scattering and has not yet been detected. Instead, the most pronounced effect of a current is a rotation of the magnetic lattice by an angle as described above.
 In Ref.~[\onlinecite{Jonietz10}] it was shown experimentally and explained 
theoretically that such a rotation 
arises from the interplay of spin-torque effects and thermal gradients in the sample.
In this paper, we want to investigate {\em other} forces which can also induce rotations of skyrmion lattices even in the {\em absence} of thermal gradients. Such forces can arise due to small distortions
of the skyrmion lattice induced by the underlying atomic lattice.
Rotations without thermal gradients have not yet been observed experimentally but may become important in future experiments and/or other materials with skyrmion phases.

In the following we will first introduce our theoretical framework based on 
the appropriate Ginzburg Landau theory (Sec.~\ref{equilibrium}) and the standard Landau-Lifshitz-Gilbert (LLG) equation (Sec.~\ref{secSTT})  for the
magnetization.\cite{slon96,berg96,zhan04} Possible modifications of the LLG
equations due to the presence of spin-orbit coupling\cite{Haney10} and its
ramifications are left for future studies. We then use the method of Thiele\cite{Thiele72}, that means we project the 
LLG equations onto the translational mode to derive an effective equation 
of motion, from which we can infer the drift velocity of the skyrmion lattice (neglecting the effects of pinning by disorder).
Extending this method, we also derive an effective equation for the rotational degree of freedom.
In section \ref{secRot} we finally apply our theory to the skyrmion lattice. As the relevant distortions of the skyrmion lattice depend sensitively on the direction of the magnetic field, we derive specific predictions for the dependence of the rotation angle on the orientation of fields and currents.

\section{Ginzburg-Landau theory for the skyrmion lattice in equilibrium}
\label{equilibrium}

We begin with a brief review of the Ginzburg-Landau theory for the
skyrmion lattice used in Ref.~[\onlinecite{mueh09}] and discuss additional terms
that orient and distort the skyrmion lattice. As we will later show, these
latter terms are necessary to enable angular momentum transfer from the
magnetization to the atomic crystal lattice. In the following, we will always
consider chiral magnets with the same symmetry (P2$_1$3) as MnSi or
Fe$_{1-x}$Co$_{x}$Si where skyrmion phases have been observed 
\cite{mueh09,muen09,yu:2010} and are predicted \cite{mueh09} to occur 
generically.

As the skyrmion lattice phase occurs only in a small temperature window close to
the classical phase transition, one can use a Ginzburg-Landau model to describe
the equilibrium properties. The weak spin-orbit coupling $\lambda_{\rm SO}$ in
MnSi gives rise to a clear separation of energy scales that allows a
classification of terms in the Ginzburg-Landau free energy in powers of
$\lambda_{\rm SO}$. The strongest energy scale is determined by ferromagnetic
exchange interactions that favor spin alignment, while the relativistic 
rotationally invariant Dzyaloshinskii-Moriya spin-orbit interaction, $D\sim
\lambda_{\text{SO}}$, favors chiral spin alignment on a weaker scale. Keeping
only terms up to order $\lambda_{\rm SO}^2$, the free energy is still
rotationally invariant,
\begin{multline}  
F[\vect {M}]=\int\!d^3r\,\Bigl( r_0\vect {M}^2+J(\nabla\vect {M})^2  \\
 + 2D\,\vect {M}\cdot(\nabla\times\vect {M}) 
 +U\vect {M}^4 -\vect {B}\cdot \vect {M}\Bigr),
\label{F} 
\end{multline}
where $\vect {M}(\vect {r})$ is the local magnetization, $\vect {B}$ the external magnetic field and $r_0,J,D,U$ are
parameters ($U,J>0$). We will choose $D>0$ that selects in the helical phase a left-handed spiral
with wavevector $Q=|\vect  Q|=D/J$. As all magnetic structures develop on the length
scale $1/Q \sim 1/D$, each gradient term contributes with a
power $\lambda_{\text{SO}}$, so that Eq.~(\ref{F}) becomes indeed of order $\lambda_{\rm SO}^2$. As we will discuss below, the magnetic structure is only oriented with respect to the atomic crystal lattice by weaker terms that are of higher order in $\lambda_{\rm SO}$ and break the rotational symmetry.\cite{naka80,bak80}

After rescaling length $\tilde{\vect {r}}=Q \vect {r}$, the magnetization $\tilde{\vect {M}}=[U/(JQ^2)]^{1/2}\vect {M}$ and field
$\tilde{\vect {B}}=[U/(JQ^2)^3]^{1/2}\vect  B$ the free energy functional reduces to
\begin{multline}
 F=\gamma\int\!d^3 \tilde r\,\left[ (1+t)\tilde{\vect  M}^2+ (\tilde\nabla
    \tilde{\vect M})^2 \right. \\
\left.  +2\,\tilde{\vect  M}\cdot(\tilde\nabla\times
    \tilde{\vect M})+\tilde{\vect M}^4 -\tilde{\vect B}\cdot \tilde{\vect
M}\right], 
\label{tF}
\end{multline}
where $\gamma=J^2Q/U$, and $t=r_0/(JQ^2)-1$ measures the distance to the
critical temperature (i.e. within the
mean-field approximation, the system is spiral spin-ordered for $t<0, B=0$ and
paramagnetic for $t>0, B=0$).
From now on, we will omit all tildes to simplify the notation, but keep in mind
that we have chosen particular units.

To describe the skyrmion lattice at finite $B$, one can expand the magnetization in plane
waves 
\begin{equation}
 \label{m} 
 \vect M(\vect
r)=\vect M_f+
\sum_{{\vect Q_j} \in L_{R}}
\left(\vect m_{\vect  Q_j}e^{i \vect Q_j\cdot\vect
r}+\mbox{c. c.}\right), 
\end{equation}
where the sum extends over all elements  ${\vect Q_j}$ of $L_R$, that denotes the reciprocal lattice
without $Q=0$, and we introduced the uniform ferromagnetic component
$\vect M_f$. The Fourier composition converges rapidly when more and more $\vect Q_j$
are included (as has been shown explicitly Ref.~[\onlinecite{mueh09SOM}]) as
the skyrmion lattice is a smooth, singularity-free texture. 
For the  rotationally invariant free energy functional Eq.~(\ref{F}),
i.e. to leading order in $\lambda_{\text{SO}}$, the reciprocal lattice 
is a two-dimensional hexagonal lattice perpendicular to the external field $\vect B$, and $\vect M_f$  is parallel 
$\vect B$. Using this ansatz in Eq.~(\ref{F}),
one obtains a local minimum of the Ginzburg-Landau free energy for a range of
parameters. This minimum is characterized by an almost constant amplitude of the
magnetization and describes in real space a lattice of skyrmions.
The skyrmion is topologically characterized by an integer winding number 
(for $\vect{B} \|\hat{ \vect z}$)
\begin{equation} \label{W}
 W=\frac{1}{4 \pi} \int\limits_{\rm UC} dx dy\,  \hat {\vect \Omega} \cdot (\partial_x \hat {\vect \Omega}
\times \partial_y \hat {\vect \Omega}),
\end{equation}
where $\hat{\vect \Omega}=\vect M/|\vect M|$ is the direction of the 
magnetization and we integrate over the magnetic unit cell (UC). As one obtains
$W=-1$, the magnetic texture corresponds to a lattice of anti-skyrmions.
As discussed in Ref.~[\onlinecite{mueh09SOM}], on the mean-field level there is
always a solution of the mean-field equations with a lower energy corresponding
to the 
''conical'' helix with a $\vect Q$ vector parallel to $\vect B$. However, for a
small range of $B$ and $t$ the mean-field energy difference between the latter
and the skyrmion lattice phase is tiny. For this reason, it is important to
consider corrections beyond the mean-field approximation. In
Ref.~[\onlinecite{mueh09SOM}] we have calculated the correction to the free energy
arising from Gaussian thermal fluctuations
around the mean-field solution which turn out to stabilize the
skyrmion phase. Due to the fluctuations the skyrmion lattice therefore
becomes a global rather than local minimum. 

\subsection{Orientation of the skyrmion lattice}
\label{Orientation}

Within the isotropic free energy functional of Eq.~\eqref{F}, the 
two-dimensional lattice spanned by the $\vect Q_j$ vectors is
always perpendicular to $\vect B$ but its orientation within this plane,
described by an angle $\Phi$, is not
fixed due to the remaining rotational symmetry around the $\vect B$ axis. This
rotational symmetry is however broken by terms of higher order in spin-orbit
coupling $\lambda_{\rm SO}$ not yet included in Eq.~\eqref{F}, which in turn
will lead to a preferential direction of $\Phi$. Due to the sixfold symmetry of
the undistorted skyrmion lattice, $\Phi$ can only be fixed by terms that 
generate an effective potential of the form $\cos(6 n \Phi-\varphi_0)$ with 
$n=1,2, ...$ and $\varphi_0 =$ const.
Lowest order perturbation theory in terms like 
$M_x^4+M_y^4+M_z^4$ or 
$(\partial^2_x \vect M)^2 + (\partial^2_y \vect M)^2 + (\partial^2_z \vect M)^2$, 
does not produce such a potential. 
One example of a term that
can lock $\Phi$ in lowest order perturbation theory is
\begin{equation}
\label{FL}
%
F_L={\gamma_L} \int d^3 r \left((\partial^3_x \vect{M})^2+(\partial^3_y \vect{M})^2+(\partial^3_z \vect{M})^2 \right).
\end{equation}
The effective potential generated by this term is extremely small because 
$\gamma_L/\gamma \sim \lambda_{\text{SO}}^4$ [note that we use the rescaled
variables of Eq.~(\ref{tF})]  
but there are no terms of lower order in 
$\lambda_{\text{SO}}$ which can orient the skyrmion lattice.

For the range of parameters considered in this work,
a positive $\gamma_L$ describes the experimental observation\cite{mueh09} 
that one of the reciprocal lattice vectors $\vect{Q}_j$
tends to be oriented in a $\langle 110\rangle$ direction (as in MnSi). A subtle
problem is the explanation of the orientation of the skyrmion lattice for a
magnetic field in $\langle 100 \rangle$ direction.
For this special case, $F_L$ of Eq.~\eqref{FL}
does not pin the angle $\Phi$ to linear order in $\gamma_L$ as $F_L$ is
symmetric under a rotation by $\pi/2$ around
$\langle 100 \rangle$ 
and $\cos[6 (\Phi+\pi/2)]=-\cos[ 6 \Phi]$. Therefore the orientation
of $\vect Q_i$ vectors is determined by effects of higher order in
$\lambda_{\rm SO}$.

\subsection{Distortion of the skyrmion lattice}

Terms of higher order in spin-orbit coupling $\lambda_{\rm SO}$ will also
distort the skyrmion lattice so that it will deviate from the perfect hexagonal
structure predicted by Eq.~\eqref{F}. In lowest order in $\lambda_{\rm SO}$,
such distortions are, for example, caused by the term
\begin{equation}
\label{FD}
F_D={\gamma_D} \int d^3 r 
\left[ (\partial_x M^y)^2 +(\partial_y M^z)^2 +(\partial_z M^x)^2 
\right]
\end{equation}
which is also written in our rescaled units. Such a term is consistent with the B20 crystal structure of MnSi. The prefactor $\gamma_D$ (in
rescaled units) is again small, $\gamma_D/\gamma \sim \lambda_{\rm
  SO}^2$, but expected to be much larger than $\gamma_L$ of Eq.~\eqref{FL},
 $\gamma_L/\gamma_D \sim \lambda_{\text{SO}}^2$. Note, however, that this term would lock
the orientation $\Phi$ only to order $\gamma_D^2 \ll \gamma_L$.

The skyrmion lattice obtained in the presence of the term \eqref{FD} is a
two-dimensional reciprocal lattice, that is a distorted hexagonal lattice. Its 
reciprocal lattice vectors $\vect Q_j$ are generically not anymore perpendicular
to the external magnetic field $\vect B$, i.e., 
$\vect B \cdot \vect Q_j \neq 0$, but to a slightly changed normal vector 
$\hat{\vect n}$. Thus, $\hat{\vect n}$ is defined by 
\begin{align} \label{normal}
\hat{\vect n} \cdot \vect Q_j = 0
\end{align}
Moreover it is normalized $\hat{\vect n}^2 = 1$ and has a positive overlap
with the magnetic field $\vect{B} \cdot \hat{\vect n} >0$. 

\section{Spin transfer and rotational torques}
\label{secSTT}

To describe the dynamics of the orientation
$\hat{\vect \Omega}(\vect{r},t)=\vect{M}(\vect{r},t)/|\vect{M}(\vect{r},t)|$ of
the magnetization $\vect{M}(\vect{r},t)$ in the presence of spin-transfer
torques due to electric currents we use the standard Landau-Lifshitz-Gilbert
(LLG) equation,\cite{slon96,berg96,zhan04} 
\begin{equation} 
\label{LLG}
\left( \partial_t + {\vect v}_s \nabla \right) {\hat{\vect \Omega}} 
= - {\hat{\vect \Omega}} \times {\vect H}_{\rm eff} 
+  \alpha \, \hat{\vect \Omega} \times \Bigl(\partial_t + \frac{\beta}{\alpha} 
{\vect v}_s \nabla\Bigr){\hat{\vect \Omega}}.
\end{equation}
Here $\vect v_s$ is an effective spin velocity parallel to the spin current
density, $\vect j_s \sim M \vect v_s
\sim \vect j_c\, p/e$, with the charge current density $\vect j_c$, the local spin
polarisation $p$ and the electron charge
$e$. 
The last two terms describe the effect of magnetization relaxation, leading to a Gilbert damping constant $\alpha$ and a dissipative spin transfer torque parameter $\beta$.
%

The magnetization precesses in the effective magnetic field 
$\vect H_{\rm eff} \approx  -\frac{1}{ M} \, \frac{\delta F}{\delta
\hat{\vect \Omega}}$.  Strictly speaking
Eq.~(\ref{LLG}) is only valid for a constant amplitude of the magnetization,
$|\vect{M}|=\text{const}$. and therefore one has
to define carefully how $\vect H_{\rm eff}$ is obtained from a Ginzburg-Landau
free energy $F[\vect M]$ when $|\vect{M}|$ is varying. As shown in Ref.~[\onlinecite{mueh09SOM}],  in the skyrmion phase the
amplitude of the magnetization varies only little so that the LLG equation,
which does not include
the dynamics of the amplitude, can be used as a good approximation. For our
numerical implementation, we use the approximation $\vect  H_{\rm eff}
\approx -\frac{1}{M} \frac{\delta F}{\delta \vect M}  \, \frac{\partial \vect
M}{ \partial
\hat{\vect \Omega}}$
where $M^2=\langle \vect M^2 \rangle$ is the average equilibrium magnetization;
other implementations $\vect  H_{\rm eff}$
will only slightly influence our results.

The LLG equation, Eq.~(\ref{LLG}), can be cast into the equivalent form
\begin{equation}
 \label{LLG1}
\hat{\vect \Omega} \times \left( \partial_t + {\vect v}_s \nabla \right) 
{\hat{\vect \Omega}}
+ {\alpha} \Bigl(\partial_t +
\frac{\beta}{\alpha} {\vect v}_s \nabla\Bigr){\hat{\vect \Omega}} 
=  {\vect H}_{\rm eff}.
\end{equation}
The skyrmion lattice of an idealized system without anisotropies 
spontaneously breaks translation and rotation invariance perpendicular to the
magnetic field. Thiele\cite{Thiele72} suggested in 1972 to project
the equations onto the relevant translational modes. We will use this
approach below and extend it in a straightforward way also to the rotational degree of freedom.
Technically, the corresponding equations of motion are obtained by multiplying
Eq.~(\ref{LLG1}) with $\hat{\vect{G}}\, {\hat{\vect \Omega}}$, where
$\hat{\vect{G}}$ is the generator of the translation or rotation mode,
and integrating over a two-dimensional unit cell (UC) of the skyrmion crystal.
With the help of these equations, we derive the drift velocity $\vect v_d$
 and the effective rotation angle $\delta \Phi$ induced by the spin-current. 

\subsection{Translational mode}

Multiplying Eq.~(\ref{LLG1}) with the generator of the translational mode, which
is given by 
$\hat{\vect G}^i_{\rm trans} \hat{\vect \Omega} 
= \partial_i \hat{\vect \Omega}$,
the occuring integral
\begin{equation}
\int d^2 r \, \bigl( \partial_i \hat{\vect \Omega} \bigr) \cdot 
\vect H_{\rm eff} 
\propto \int d^2 r \, \bigl(\partial_i \hat{\vect \Omega}\bigr) \cdot 
\frac{\delta F}{\delta\hat{ \vect \Omega}} 
\end{equation}
vanishes due to translational
invariance. In the stationary limit, where the magnetic structure drifts with a constant velocity,
 $\hat{\vect \Omega}=\hat{\vect \Omega}(\vect r-\vect v_d t)$,
one obtains \cite{Thiele72,he06} 
\begin{equation} 
\label{drift1}
\vect G \times  \left( {\vect v}_s-{\vect v}_d \right) + \vect{\mathcal D}
\left(\beta{\vect v}_s-\alpha {\vect v}_d  \right) = 0
\end{equation}
with
\begin{subequations}
\begin{align}
\vect{G}_i &=\epsilon_{ijk} \frac{1}{2} \int_{\rm UC} d^2 r\, 
\hat{\vect \Omega} \Bigl(\partial_j {\hat{\vect \Omega}} \times
\partial_k {\hat{\vect \Omega}} \Bigr) 
\\
\vect{\mathcal{ D}}_{ij} &=\int_{\rm UC} d^2 r\, \partial_j 
{\hat{\vect\Omega}}\,
\partial_i {\hat{\vect \Omega}} 
\end{align}
\end{subequations}
In Ref.~[\onlinecite{Thiele72}], the vector $\vect G$ (up to prefactors) has been
identified by Thiele as a ``gyrocoupling
vector'' as it translates a spin current to an effective Magnus force in
perpendicular direction. In fact, the gyrocoupling vector is just proportional
to the winding number $W$ of Eq.~\eqref{W} per unit cell,
\begin{align} \label{GyroT}
\vect{G}_i = 4 \pi W \hat{\vect {n}}_i,
\end{align}
and points in a direction orthogonal to the two-dimensional skyrmion
lattice, see Eq.~\eqref{normal}. It is therefore topologically
quantized . Note that in the presence of a distortion term like
Eq.~\eqref{FD}, the surface normal $\hat{\vect n}$ is not necessarily
parallel to the applied magnetic field $\vect B$. The effective Magnus
force caused by $\vect G$ must be equal to a counter force to the
electrons. Indeed, electrons which follow adiabatically the magnetic
texture, pick up a geometric Berry phase which results in a transverse
force corresponding to an effective magnetic field of strength $|\vect
G|/(4 \pi A_{\rm UC})$ where $A_{\rm UC}$ is the area of the magnetic
unit cell.  The resulting topological Hall effect of the expected
strength has been already observed in Ref.~[\onlinecite{neub09}].

The dimensionless matrix $\vect{\mathcal D}$ is called the
`dissipative tensor'\cite{Thiele72} as it describes together with
$\alpha$ and $\beta$ the effects of
dissipative forces on the moving skyrmion lattice. As the magnetization direction $\vect\Omega$
only varies within the two-dimensional plane of the skyrmion lattice, the
$3\times 3$ matrix $\vect{\mathcal D}$ possesses a zero eigenvalue corresponding
to the normal direction $\hat{\vect n}$. Within the plane of the skyrmion
lattice, the matrix $\vect{\mathcal D}$ is diagonal in lowest order in
spin-orbit coupling $\lambda_{\rm SO}$ due to the 6-fold symmetry of the
skyrmion lattice. So we can approximate to lowest order in $\lambda_{\rm SO}$
\begin{align} \label{DissTapp}
\vect{D}_{ij} \approx \mathcal D \vect P_{ij}
\end{align}
with the projector $\vect P_{ij} = (\mathds{1} - \hat{\vect n}\cdot
\hat{\vect{n}}^T)_{ij}$. To lowest order in the current $\vect v_s$, $\vect G$
and $\vect {\mathcal D}$ can be evaluated using
the equilibrium magnetization.

For most magnetic bulk structures $\vect G$ vanishes and therefore 
$\vect v_d=\vect v_s \beta/\alpha$. For the topological non-trivial skyrmion
lattice phase, we have instead a finite skyrmion density $W = -1$ and thus a
finite $\vect G$. From Eq.~\eqref{drift1} we then obtain for the in-plane drift
velocity $\vect  v_d^\parallel = \vect P \vect v_d$ in agreement with
Ref.~[\onlinecite{he06}]
\begin{eqnarray}
{ \vect  v_d^\parallel}&=&\frac{\beta}{\alpha} { \vect v_s^\parallel}+ \frac{\alpha-\beta}{\alpha^3
(\mathcal D/4 \pi W)^2+\alpha} \Bigl( { \vect
v_s^\parallel} - \frac{\alpha \mathcal D}{4 \pi W}\hat{\vect{n}}
\times {\vect v_s^\parallel}\Bigr)\nonumber \\
&\approx&
\vect{v}_s^\parallel-\frac{(\beta-\alpha)\mathcal D}{4 \pi W}
\hat{\vect n} \times \vect{v}_s^\parallel  
\label{vd}
\end{eqnarray}
with the in-plane spin-velocity $\vect  v_s^\parallel = \vect P \vect
v_s$  and $\hat{\vect n}$ is the normal vector given in 
Eq.~\eqref{normal}. The last line is obtained in the limit $\alpha,\beta\ll 1$, 
which is appropriate in the limit of small $\lambda_{\rm SO}$ as the damping
terms arise from spin-orbit coupling effects. 
The drift velocity $\vect v_d^\parallel$ of the skyrmion lattice is not parallel
to the spin velocity $\vect v_s^\parallel$ 
due to the Magnus forces arising as counter forces to the topological Hall
effect.\cite{neub09}

\subsection{Rotational mode}
\label{RotationalMode}

The rotational mode differs in several aspects from the translational one.
Most importantly, weak spin-orbit interactions break rotational invariance and
therefore a rotational torque due to the current can be balanced by a counter
torque of the underlying atomic crystal lattice.
Note that an infinitely large skyrmion domain formally needs an infinite time 
to reorient due to the fact that a rotation by a small angle leads to infinitely
large time-dependent changes of the magnetization at large distances and to 
dissipation forces. In practice, domains are always finite and we will therefore
proceed calculating the change of the steady-state orientation of the skyrmion
lattice in the presence of a small current.

In order to derive the equation of motion for the rotational degree of freedom we need the generator of rotations, $\hat{\vect{G}}_{\rm rot}$. The magnetic texture rotated by a finite angle $\phi$ around the axis defined by the normal vector $\hat{\vect n}$ of Eq.~\eqref{normal} is generally given by $\hat{\vect \Omega}'(\vect r) = R_{\hat{\vect n}}(\phi) \hat{\vect \Omega}(R^{-1}_{\hat{\vect n}}(\phi) \vect r)$ where $R_{\hat{\vect n}}(\phi)$ is the rotation matrix. For infinitesimal angles this rotation matrix reads $(R_{\hat{\vect n}}(\phi))_{ij} = \delta_{ij} + \phi \epsilon_{ikj} \hat{\vect n}_k + \mathcal{O}(\phi^2)$ with the Levi-Civita tensor $\epsilon_{ikj}$. From this follows the needed generator $\hat{\vect \Omega}'(\vect r) - \hat{\vect \Omega}(\vect r) = \phi \hat{\vect{G}}_{\rm rot} \hat{\vect\Omega}  + \mathcal{O}(\phi^2)$ with  $\hat{\vect{G}}_{\rm rot} \hat{\vect\Omega} = \hat{\vect{n}} \times
\hat{\vect \Omega} - (\hat{\vect n} ({\vect r} \times \nabla) )
\hat{\vect \Omega}$.
As above, we multiply Eq.~(\ref{LLG1}) by 
$\hat{\vect{G}}_{\rm rot} \hat{\vect\Omega}$ and integrate over space to obtain
an effective equation for the rotational degree of freedom
\begin{equation} 
\label{Torque1}
{\vect P}_{R} \left( {\vect v}_s-{\vect v}_d \right) + {\vect P}_{D} \left(\beta{\vect
v}_s-\alpha {\vect v}_d  \right) =
\tau
\end{equation}
The left-hand side describes how a rotational torque is created by the applied
current. In general, a matrix is needed to describe the current induced
rotational motion. As we consider only rotations around a given axis, we can
instead use the two vectors $\vect P_R$ and $\vect P_D$,
\begin{subequations}
\label{Pvectors}
\begin{align}
{\vect P}^i_{R} &=\int_{\rm UC} d^2 r\, \left( \hat{\vect \Omega}
\times \partial_i \hat{\vect \Omega} \right) 
(\hat{\vect{G}}_{\rm rot} {\hat{\vect \Omega}}), \label{pr}
\\
{\vect P}^i_{D} &=\int_{\rm UC} d^2 r\, \partial_i \hat{\vect \Omega}
\,(\hat{\vect{G}}_{\rm rot} {\hat{\vect \Omega}}), \label{pd}
\end{align}
\end{subequations}
 which we term the
reactive and dissipative rotational coupling vectors, respectively. These 
central quantities describe how a velocity leads to a torque around the axis
defined by the normal $\hat{\vect n}$.
The reactive term arises from the Berry phases picked up by spin currents in the
presence of a non-trivial spin-texture while the dissipative terms can directly
be traced back to the damping terms. As the magnetic texture in the skyrmion
lattice phase only varies within a plane, the two coupling vectors are
orthogonal to its normal, $\vect P_R \cdot \hat{\vect n} = 0$ and $\vect P_D
\cdot \hat{\vect n} = 0$. As a consequence, only the in-plane velocities $\vect
v_d^\parallel$ and $\vect v_s^\parallel$ enter Eq.~\eqref{Torque1}.

The rotational torque exerted by the current is balanced on the
right-hand side of Eq.~\eqref{Torque1} by 
the flow of angular momentum from the skyrmion lattice to the
underlying atomic lattice. The torque
\begin{align} \label{LatticeTorque}
\tau &= \int_{\rm UC} d^2 r\, {\bf H}_{\rm eff} (\hat{\vect{G}}_{\rm rot} {\hat{\vect \Omega}})=-\frac{\partial f}{\partial \Phi}\approx -\chi\, \delta \Phi
\end{align}
can be expressed by the change of free energy (per magnetic unit cell
and divided by the magnetization) upon a rotation of the
magnetic structure by the angle $\Phi$. To obtain the correct sign
note that $\Phi$ describes the rotation of the magnetization and not
of the coordinate system.
In the linear response regime, the torque $\tau$ can be expanded in the small
deviation $\delta \Phi$ from the equilibrium orientation, see discussion in
section \ref{Orientation}. The restoring force depends on the susceptibility,
i.e., the ``spring constant"
\begin{align} \label{Susc}
\chi = \frac{\partial^2 f}{\partial \Phi^2}.
\end{align}
In Eq.~\eqref{LatticeTorque}, we further used that the torque $\partial
f/\partial \Phi$ vanishes in equilibrium for $\vect v_s=0$.

When discussing Eqs.~(\ref{Pvectors}) a careful interpretation of the terms linear in
$\vect{r}$ are necessary which arise
for any rotational mode. We have checked 
 that they vanish for symmetrically
shaped macroscopic domains. These terms give, however, extra rotational torques for domains
with asymmetric shape with shape-dependent sign and strength. 
Assuming that these average to zero, we neglect all terms linear in
$\bf r$ and approximate
\begin{subequations}
\begin{align}
{\vect P}^i_{R} &\approx \int_{\rm UC} d^2 r\, \left( \hat{\vect \Omega}
\times \partial_i \hat{\vect \Omega} \right) 
\left(\hat{\vect{n}} \times \hat{\vect \Omega}\right),
\label{prappr}
\\
{\vect P}^i_{D} &\approx \int_{\rm UC} d^2 r\, \partial_i \hat{\vect \Omega}
\,\left(\hat{\vect{n}} \times \hat{\vect \Omega}\right).
\label{pdappr}
\end{align}
\end{subequations}

Within linear response, we can solve Eq.~\eqref{Torque1} for the rotational
angle $\delta \Phi$, (provided that $|\chi| \neq 0$)
\begin{align}
\label{delPhi}
\delta \Phi = - \frac{1}{\chi} \left [ {\vect P}_{R} \left( {\vect v}_s-{\vect v}_d \right) + {\vect P}_{D} \left(\beta{\vect
v}_s-\alpha {\vect v}_d  \right) \right],
\end{align}
where in linear order in $\vect v_s$ the coefficients are again evaluated with
the equilibrium magnetization.
Together with the equation for the drift velocity, Eq.~(\ref{vd}), this is our central result (together with its numerical evaluation discussed below).
Note that only the in-plane velocities $\vect v_s^\parallel$ and
$\vect v_d^\parallel$ enter Eq.~\eqref{delPhi} and generate a rotation as the
orthogonal components are projected out. Using the explicit solution for the
drift velocity, Eq.~(\ref{vd}), the equation for $\delta \Phi$ simplifies for
small $\alpha$ and $\beta$ to
\begin{equation} 
\label{delPhiApprox}
\delta \Phi \approx - \frac{\beta-\alpha}{\chi}
\Bigl(\frac{\mathcal D}{4 \pi W}  {\vect P}_{R } (\hat{\vect{n}} \times \vect v_s) 
+ {\vect P}_{D } \vect v_s  \Bigr)
\end{equation}
with $W=-1$ for the Skymion lattice.
As many other spin-torque effects,
\cite{Tserkovnyak08} the rotation vanishes for $\alpha=\beta$ where the effective
Galileian invariance of Eq.~(\ref{LLG}) allows only for trivial solutions where
the magnetic structure drifts with the current.

\section{Theory for $\delta \Phi$ in skyrmion lattices}
\label{secRot}

For the undistorted skyrmion lattice, i.e. for the rotationally symmetric
Ginzburg-Landau free energy Eq.~(\ref{F}), the rotation angle $\delta \Phi$
vanishes by symmetry. 
The sixfold rotational symmetry of the hexagonal magnetic lattice 
implies immediately that the two rotational coupling vectors $\vect{P}_R$ and
$\vect{P}_D$ of Eqs.~\eqref{Pvectors} have to vanish.
More precisely, as the orientation of a hexagon is described by a third-rank
tensor, a rotational torque will only show up to order $\vect v_s^3$, too small
for any observable effect (at least in bulk materials, where only
relatively small current densities can be applied, see Ref.~[\onlinecite{Jonietz10}]). To obtain an effect already in linear order in $\vect
v_s$, 
one has to take into account that the skyrmion lattice is slightly
distorted by a coupling to the underlying atomic crystal
lattice. Furthermore, one has to investigate the origin of the
restoring forces, $\chi \delta
\Phi$, which also arise from higher-order spin-orbit coupling terms.

We have systematically investigated which symmetry-allowed terms in the
Ginzburg-Landau theory to leading order in $\lambda_{\text{SO}}$ give rise to
$(i)$ a distortion allowing for
rotational coupling or $(ii)$ a preferred orientation of the magnetic lattice.
Two representatives
of such terms are given by $F_L$ in Eq.~(\ref{FL}) and $F_D$ in Eq.~\eqref{FD}  
with coupling constants $\gamma_L$  and $\gamma_D$, repectively. 
$F_D$ leads to small but finite rotational coupling vectors, such that 
$\vect P_R$ and $\vect P_D$ are of order $\mathcal{O}(\gamma_D)$.
The term $F_L$, on the other hand, orients the skyrmion lattice and gives rise
to a finite susceptibility $\chi \sim \mathcal{O}(\gamma_L)$. While $\gamma_D$
is very small, a sizable $\delta \Phi$ is only obtained
because the susceptibility $\chi$ is even smaller $\gamma_L/\gamma_D \sim
\lambda_{\text{SO}}^2$.  
As a consequence, a sizable effect $\delta \Phi \propto \gamma_D/\gamma_L$ can
be expected.

\subsection{Numerical solution of  the skyrmion lattice}

For the numerical evaluation of $\delta \Phi$, we have to evaluate the 
coefficients of the formula for the rotation angle, Eq.~\eqref{delPhi}, with the
equilibrium magnetization texture $\vect M(\vect r)$ of the skyrmion lattice
obtained from the free energy functional of Eq.~\eqref{tF} together with
Eqs.~\eqref{FL} and \eqref{FD}. In order to obtain $\vect M(\vect r)$, we employ
the following mean-field approximation. We minimize the free energy functional
with the ansatz of Eq.~\eqref{m} for the magnetization but we include in the sum
only the three smallest reciprocal lattice vectors $\vect Q_j$. This is a good
approximation as it turns out that higher order terms contribute (both
experimentally and theoretically\cite{mueh09}) maximally a few
percent to the total magnetization.
For simplicity, we do not include the effects of thermal fluctuations in our
calculation. While these are essential to promote the skyrmion
solution from a local minimum to a global minimum of the free energy,
they are expected to give rise only to small renormalization of
prefactors at least not too close to the phase transition as was checked
explicitly in Ref.~[\onlinecite{mueh09SOM}].

Hence, we will approximate the static magnetic texture in total by 27 real
parameters $\mu_i$,
$i=1,...,27$: the uniform magnetization $\vect M_f$, two reciprocal lattice
vectors $\vect{Q}_1$, $\vect{Q}_2$
(with $\vect{Q}_3=-\vect{Q}_1-\vect{Q}_2$) and three complex vectors $\vect
M_{\vect{Q}_j}$, $j=1,2,3$. In this representation, 
the solution that accounts for the sliding motion of the skyrmion lattice 
can then be written in the absence of pinning forces by disorder and to linear order in the applied current as 
\begin{equation}
\vect{M}(\vect r,t)=\vect{M}(\vect r - \vect v_d t,  \{\mu_i\})
\label{magnetvar} 
\end{equation}
where  $\vect{v}_d$ is the drift velocity of
the magnetic structure.

\subsection{Numerical evaluation of $\delta \Phi$}

With the help of the magnetic texture $\vect M(\vect r)$ in equilibrium we can 
evaluate the gyrocoupling vector $\vect G$, the dissipative tensor 
$\vect{\mathcal D}$, the rotational coupling vectors $\vect P_D$ and 
$\vect P_L$, and the susceptibility $\chi$ that are needed to determine the 
rotational angle $\delta \Phi$ using Eq.~\eqref{delPhi}. In the limit of small 
$\alpha,\beta \ll 1$, Eq.~\eqref{delPhi} reduces to Eq.~\eqref{delPhiApprox}, 
and  after collecting all prefactors arising from the rescaling of variables, 
we then obtain
\begin{equation} \label{DelPhiApprox2}
 \delta \Phi = \Phi_0
\, \delta \varphi(t,\tilde {\vect  B}, \hat{\vect j})
\quad {\rm for}\quad  \alpha,\beta,\gamma_L,\gamma_D \ll 1,
\end{equation}
where $\Phi_0$ is given by 
\begin{align} \label{Phi0}
\Phi_0 = v_s \frac{\hbar  ({\alpha}- {\beta})  \sqrt{U} }{Q^2 J^{3/2}} \,\frac{\gamma_D}{ \gamma_L}.
\end{align}
The dimensionless function $\delta \varphi(t, \tilde {\vect B}, \hat{\vect j})$
depends only on the dimensionless distance
from the critical point $t$, the direction and strength of the dimensionless
magnetic field $\tilde {\vect
B}$, see Eq.~\eqref{tF}, and the orientation of the current.

We will first discuss the dependence of $\delta \Phi$ on the relative and
absolute orientation of the magnetic field $\vect B$ and applied current $\vect
v_s$. Afterwards we discuss the dependence of $\delta \Phi$ on the distance,
$t$, to the phase transition.

\subsubsection{Orientational dependence of $\delta \Phi$}

\begin{figure}[t]
\begin{center}
\includegraphics[width=8.5cm,clip]{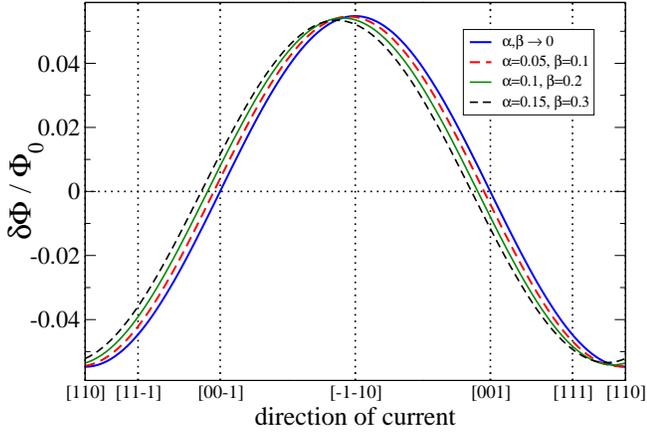}
\end{center}
\caption{Effective rotation angle $\delta \Phi$ in units of $\Phi_0$, 
Eq.~\eqref{Phi0}, for a magnetic field, $\vect B$, in [1-10] direction as a
function of the current direction perpendicular to [1-10] for various values of
$\alpha,\beta$ given in the inset. The other chosen parameters are 
$t=-0.8$, $|\tilde {\vect B}|=0.5 \sqrt{-2 t}$,
$\gamma_D=0.01$ and $\gamma_L=0.001$.}
\label{fig:current}
\end{figure}

\begin{figure}[t]
\begin{center}
\includegraphics[width=8.5cm,clip]{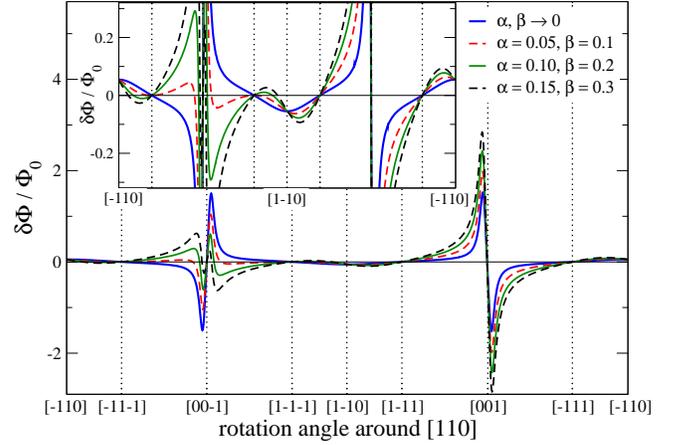}
\end{center}
\caption{Effective rotation angle $\delta \Phi$ in units of $\Phi_0$, 
Eq.~\eqref{Phi0}, for a current, $\vect v_s$, in the [110] direction
as a function of the direction of the magnetic field, $\vect B$, perpendicular 
to [110] (other parameters as in Fig.~\ref{fig:current}).
Note that for the high symmetry directions $\langle 111 \rangle$ and 
$\langle 100 \rangle$ $\delta \Phi$ vanishes.
Close to the $\langle 100 \rangle$ direction the effect is maximal (see text).
For certain directions, the effect depends sensitively on the size of $\alpha$ (in the rescaled units
$\beta$ is less important) as it affects the direction of the drifting skyrmion lattice.
\label{rot110.fig}}
\end{figure}

Within our theory, the rotation angle $\delta \Phi$ is
proportional to the product of $\vect v_s$ and a vector in the skyrmion lattice 
plane, which is almost perpendicular to the magnetic field $\vect B$ with 
deviations of order $\gamma_D$.  As a consequence, when rotating the current around the direction of a magnetic field, a simple cosine dependence is obtained. This is demonstrated in Fig.~\ref{fig:current} where a numerical evaluation of $\delta \Phi$ using Eq.~\eqref{delPhi} as a function of $\vect v_s$ for various values of $\alpha$ and $\beta$ is shown. The blue solid line corresponds to the limit of small 
$\alpha, \beta$ where $\delta \Phi$ reduces to Eq.~\eqref{DelPhiApprox2}. 

\begin{figure}[t]
\begin{center}
\includegraphics[width=8.5cm,clip]{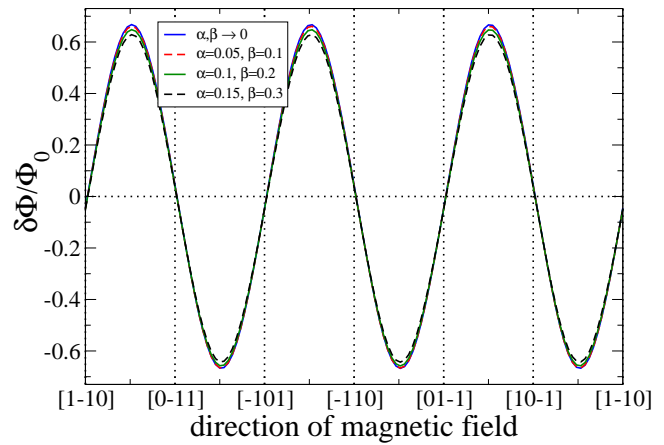}
\end{center}
\caption{Effective rotation angle $\delta \Phi$ in units of $\Phi_0$, Eq.~\eqref{Phi0}, for a current, $\vect v_s$, in the [111] direction
as a function of the direction of the magnetic field, $\vect B$, perpendicular to [111] (other parameters as in Fig.~\ref{fig:current}).\label{rot111.fig}}
\end{figure}

\begin{figure}[t]
\begin{center}
\includegraphics[width=8.5cm,clip]{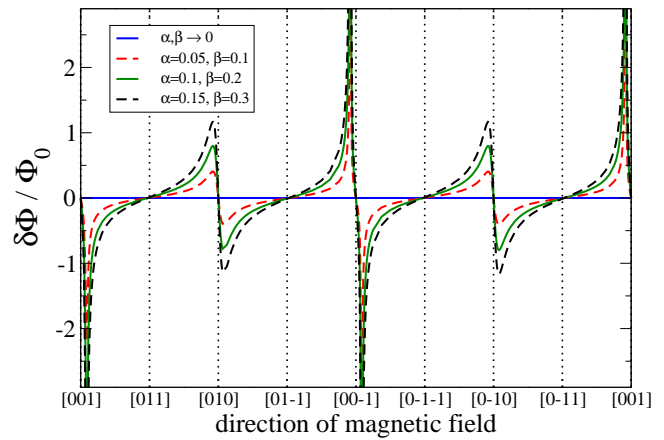}
\end{center}
\caption{Effective rotation angle $\delta \Phi$ in units of $\Phi_0$, Eq.~\eqref{Phi0}, for current, $\vect v_s$,  in the [100] direction
as the function of the direction of the magnetic field, $\vect B$, perpendicular to [100] (other parameters as in Fig.~\ref{fig:current}). Note that for this orientation, the rotation angle vanishes quadratically in the limit $\alpha,\beta \to 0$ due to a special symmetry, see text. 
\label{rot100.fig}}
\end{figure}

The dependence of $\delta \Phi$ on the direction of the field, $\hat{\vect B}$,
on the other hand, is substantially more
complicated. The three figures, Fig.~\ref{rot110.fig}, Fig.~\ref{rot111.fig} and Fig.~\ref{rot100.fig}, give an overview on how $\delta \Phi$
depends on the orientation of $\hat{\vect B}$ for fixed current direction. Several of the main features in these figures can be understood from symmetry considerations as explained in the following. Special properties can not only be expected when the field is oriented 
along either a two-fold  $\langle 100 \rangle$ axis or a three-fold  $\langle 111 \rangle$ axis, but also for a field perpendicular to
a $\langle 100 \rangle$ axis as in this case the product of time-reversal and  $\pi$ rotations around $\langle 100 \rangle$
maps $\bf B$ (and the skyrmion lattice) upon itself.

Fig.~\ref{rot110.fig} shows how the rotation angle changes as the field
$\vect B$ is rotated around the [110] axis for a current parallel to
$[110]$. For this geometry, one of the reciprocal lattice vectors is
in the absence of a current always oriented in the
[1-10]
direction for the parameters chosen in Fig.~\ref{rot110.fig}.
Note that the directional dependence is {\rm not} very universal
and depends on the structure, sign and size of the dominant anisotropy
terms and even the size of the damping constants. A
universal feature is, however, that $\delta \Phi$ vanishes (to linear order in $\vect v_s$) for magnetic fields in one of the two relevant high-symmetry direction (note that 
$\langle 110 \rangle$ is {\em not} a symmetry axis of the B20 
structure of MnSi). As $\langle 100 \rangle$ is a two-fold screw axis and $\langle 111 \rangle$
a three-fold symmetry axis, both $\vect{P}_D$ and $\vect{P}_R$ and, consequently,
also $\delta \Phi$ are zero for a magnetic field oriented in these
directions (as $\vect{P}_D$ and $\vect{P}_R$ are
perpendicular to $\bf B$ because $\hat n$ becomes parallel to $\bf B$
for these high-symmetry directions).
Away from the two high-symmetry directions, $\langle 100 \rangle$ or $\langle
111 \rangle$, we obtain finite coupling vectors, $\vect{P}_D$ and
$\vect{P}_R$, and a transfer of
angular momentum from the magnetic texture to the crystal lattice resulting in a finite
$\delta \Phi$.

As we have limited our analysis to effects that are linear in the current, a
reversal of the
current direction always leads to a sign change of the rotation angle, $\delta
\Phi \to - \delta \Phi$. To linear order in the damping coefficients $\alpha$
and $\beta$, where Eq.~\eqref{delPhi} reduces to Eq.~\eqref{delPhiApprox}, also
a reversal of the magnetic field
direction, $\vect B \to -\vect B$
 leads to a sign reversal $\delta \Phi \to - \delta \Phi$ (blue solid line in
Fig.~\ref{rot110.fig}). This is {\em not} the case if contribution of higher order
in $\alpha, \beta$ are incorporated (time reversal is broken not only
by $\bf B$ but also by the applied current and dissipative and
reactive forces have opposite signatures under time reversal). 
Even for realistic values of $\alpha \sim 0.1$,\cite{date77} the latter can have a large
effect and can even change the
 sign of $\delta \Phi$ for certain crystallographic directions, see 
Fig.~\ref{rot110.fig}.
 As a consequence, there is generically  no specific symmetry with respect to a
reversal of the current and the magnetic field,
 $\delta \Phi(\vect B,\vect j)\neq \delta \Phi(-\vect B,-\vect j)$. 

A geometry,
where the field is oriented along the $[ 1-10 ]$ direction and the current along the $[110]$ direction, is an
exception as 
$\langle 100 \rangle$ 
is a two-fold rotation axis 
which allows to map $\vect{B} \to -\vect B$ and $\vect j \to -\vect
j$. Precisely this geometry has been studied experimentally in
Ref.~[\onlinecite{Jonietz10}], where the {\em same} scattering pattern
was observed when both field and current were reversed. Within our
conventions for $\delta \Phi$ (defined relative to the field
orientation), this corresponds to a reversal of $\delta \Phi$ when
both current- and field direction are reversed. In contrast, our symmetry
analysis above shows the opposite behavior (for all symmetry-allowed anisotropies
and even beyond linear response theory). This discrepancy was resolved
in Ref.~[\onlinecite{Jonietz10}] where it was shown that additional symmetry
breaking temperature gradients explain the experiments, see introduction.

The specific directions $\langle 100 \rangle$ for the magnetic field require extra consideration as for this
orientation the susceptibility $\chi$ of Eq.~\eqref{Susc} vanishes to linear
order in $\gamma_L$ as discussed below Eq.~(\ref{FL}). More precisely, denoting by $\delta$
the angle between $\hat{\vect B}$ and 
$[001]$
(for the geometry
of Fig.~\ref{rot110.fig}), the relevant potential is proportional to
$\gamma_L \delta^2 \cos 6 \Phi$ while $\vect P_R$ and $\vect P_D$ vanish {\em
linearly} in $\delta$. Therefore
$\delta \Phi \sim 1/\delta$ as can be seen in Fig.~\ref{rot110.fig}. Only for very
small $\delta$ effects of order
$\gamma_L^2$ lead to a rounding of the divergence. If $\hat{\vect B}$ is
precisely oriented in 
[001],
one also has to include further
anisotropy terms in the analysis,
see the discussion in section \ref{Orientation}, and the predictive power of 
our theory along such special symmetry directions is limited.

In Fig.~\ref{rot111.fig} the current-induced rotation angle is shown for a current in
the $[111]$ direction when the magnetic field is rotated perpendicular
to this direction. By symmetry the pattern repeats itself every
$120^{\rm o}$. Note that $\delta \Phi$ becomes small but does not
vanish for fields in the $\langle 110 \rangle$ directions, see also Fig.~\ref{fig:current}.

For current in the $[100]$ direction and magnetic field perpendicular
to $[100]$ one again obtains large values for $\delta \Phi$ when the
field points in a $\langle 001 \rangle$ direction. Remarkably, $\delta
\Phi/\Phi_0$ vanishes exactly in the limit $\alpha,\beta \to 0$ for
this configuration while it is finite for other orientations of $\bf
B$ and $j$. The reason is again
a special symmetry: the product of time reversal $T$ and a rotation by
180$^o$ around the $[100]$ direction. Under this symmetry, $\bf B$ is
mapped upon itself. It also enforces that the reactive
rotational coupling vector, ${\bf P}_R$, which is even under $T$, points in $[100]$
direction while the dissipative rotational coupling vector, ${\bf
  P}_D$, which is odd under $T$, has to be perpendicular to
$[100]$. Using Eqs.~(\ref{vd})  one obtains that
${\vect v}_s-{\vect v}_d$ becomes perpendicular to $\vect v_s$ while $\beta {\vect
  v}_s-\alpha {\vect v}_d$ becomes parallel to $\vect v_s$ to leading order
in $\beta$ and $\alpha$. Therefore all current induced torques vanish
for $\alpha,\beta \to 0$ according to (\ref{Torque1}).

\subsubsection{Dependence of $\delta \Phi$ on the distance to the phase
transition and order of magnitude estimate}

When discussing the physics close to the phase transition, one first has to emphasize that all phase transitions are expected to be of first order. This is obvious already on the mean field level both for the transition to the conical phase (where the ordering vectors jump)
and for the transition to the paramagnetic phase (due to the presence of a cubic term in the Ginzburg Landau description in the presence of a finite magnetic field). The latter transition, is, however, also strongly affected by thermal fluctuations which, drive even the transition from the paramagnetic to the helical phase first order.\cite{bak80} This effect makes it very difficult to estimate the precise location of
the phase transition line.

As all transitions are first order, the rotation angle is formally non-singular at the transition. In practice, however, a complex interplay of phase transition dynamics, pinning effects, the external drive by currents, heating effects and even surface properties can be expected in the regime where both phases are locally stable. These questions are certainly far beyond the goal of the present study.

\begin{figure}[t]
\begin{center}
\includegraphics[width=8cm]{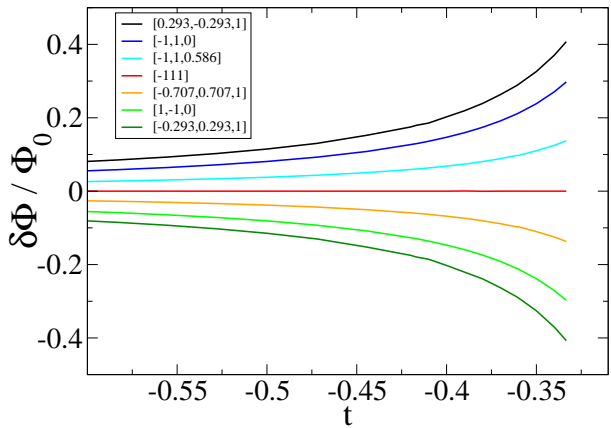}
\end{center}
\caption{Angle $\delta \Phi$ 
in the limit of small $\alpha, \beta$, see Eq.~\eqref{DelPhiApprox2},
for a current in [110] direction and various orientations of the magnetic field,
see legend, as a function of the dimensionless distance $t$ 
from the critical point ($t=-0.8$, $|\tilde {\vect B}|=0.5 \sqrt{-2 t}$,
$\gamma_D=0.01$ and $\gamma_L=0.001$). \label{rotT.fig}}
\end{figure}

We therefore restrict our analysis to the overall dependence of  the rotation angle on the parameter $t$, which can be controlled by temperature in the experiments.  Fig.~\ref{rotT.fig} displays
the dependence of $\delta \Phi$ on $t$ for fixed
magnetic field strength and various orientation of $\vect B$ in the limit of small $\alpha$ and $\beta$ when  
Eq.~\eqref{DelPhiApprox2} is valid. For the
chosen set of parameters, $\delta \Phi$ grows upon increasing $t$, i.e. for growing temperature. The overall magnitude of the effect, might, however, be overestimated as the mean-field theory cannot describe the first-order transitions quantitatively. 
Qualitatively different results for  the $t$ dependence are obtained in cases where anisotropy terms dominate  which  are not quadratic
in the order parameter (as assumed by us) but are of higher order implying a larger sensitivity to the distance to the phase transition.

Finally, we would like to estimate the order of magnitude of the rotation.
As both $\gamma_L$ and $\gamma_D$ are unknown, it is not possible to predict
quantitatively
 the size of the expected rotation angle $\delta \Phi$. One can, however, make a
crude order of
magnitude estimate by counting powers of spin-orbit coupling
$\lambda_{\text{SO}}$.
For $ t \sim 1$  we can estimate $\delta \Phi \sim \Phi_0$, see
Eq.~\eqref{Phi0}. We approximate $Q=\lambda_{\rm SO} /a$,  $a^3 \sqrt{U}/J^{3/2}\sim 1/k_B T_c$, $\gamma_{L}/\gamma_D \sim \lambda_{\rm SO}^2$ (see above) and with the drift velocity of charge, $v_s \sim j a^3$ to obtain
\begin{align}
\delta \Phi  \sim
\frac{\hbar j  a^2}{e k_B T_c}  \frac{\alpha-\beta}{\lambda^{4}_{\rm SO}}
\end{align}

For MnSi spin torque effects were observed \cite{Jonietz10} for currents of the order of $10^6\,$A/m$^2$. In the appropriate dimensionless units this corresponds to 
$\frac{\hbar j  a^2}{e k_B T_c}\sim 10^{-7}$ (using the lattice constant $a\approx 4.2$\AA\ and $T_c \approx 30$K), which shows the smallness of the applied currents (in most spin-torque experiments currents are 5-6 orders of magnitude larger). 
In MnSi $\alpha\sim 0.1$ appears to be surprisingly large as
electron-spin-resonance experiments show a rather broad peak.\cite{date77}
Using $\lambda_{\text{SO}} \sim  0.01$, one could in principle obtain sizable
rotation angles 
$\delta \Phi \sim \mathcal{O}(1)$. Experimentally, no such rotation was observed
in an experimental setup which avoids temperature gradients. Taking the
crudeness of the estimates given above into account, this result is
unfortunately also consistent with our
analysis, especially as for the experimental setup (field along $[110]$, current along [1-10]) the effect turns out to be suppressed
by another factor of $0.05$, see Fig. \ref{fig:current}.

\section{Summary and discussion}
\label{sec:summary}

The focus of this paper is the investigation of a specific mechanism how spin transfer torques can lead to a spatial rotation
of the magnetic skyrmion texture by a finite angle.  Our analysis started from 
the observation that a perfect skyrmion lattice
has neither a preferred orientation perpendicular to the applied magnetic field nor can a small current exert (to linear order) a rotational torque to such a symmetric structure  due to its sixfold rotational symmetry. The magnetic texture is, however, embedded in the atomic crystal of the host
material. This environment breaks rotational symmetry leading both to a
preferred orientation of the skyrmion lattice in equilibrium by tiny spin-orbit
coupling effects and also to a small distortion of the skyrmion lattice which enables the current to exert rotational torques. The balance of these two effects determines the rotation angle.  A systematic analysis of such effects is possible as all relevant phenomena are controlled by weak spin-orbit coupling effects and occur on long length scales.
 Overall the following picture emerges:
First, angular
momentum is transfered from the spins of the conduction electrons to the
magnetization. This induces a rotation until all rotational torques are balanced and the angular momentum 
is flowing from the magnetic texture (via spin-orbit coupling effects) to the underlying atomic lattice.

There are several other effects which can also lead to rotational torques. Most
relevant for the experiment in Ref.~[\onlinecite{Jonietz10}]
is that macroscopic inhomogeneities can lead to inhomogeneous forces and
therefore also to rotational forces. In Ref.~[\onlinecite{Jonietz10}]
these could be controlled experimentally by small temperature gradients in the
sample. A temperature gradient leads to a different strength of forces at the
`hot' and the `cold' side of a magnetic domain and therefore to torques.
Reversing the temperature gradient therefore leads to a reversal of the rotation
angle. Moreover the shape of a magnetic domain can be the origin of forces which
orient the domain in the presence of a current. These torques are likely to be
of random sign and might be responsible for the smearing of the neutron
scattering signal observed in Ref.~[\onlinecite{Jonietz10}]. Finally, also
distortions of the skyrmion lattice by disorder can lead to a reorientation of a
sliding lattice. This physics, which is based on the a non-linear response of
the moving lattice, has previously been investigated for vortex lattices in
superconductors.\cite{schm73,forgan:PRL2000} Finally, also the current itself
can distort the skyrmion lattice and lead to a reorientation of the lattice. By
symmetry, this effect occurs, however, only to third order in the current
density.

Besides the more widely studied translational motion, we expect that also the 
rotation of magnetic textures will continue to be an important signature of spin-torque 
effects.\cite{Pribiag:2007p8556} Depending on the setup 
both rotation and translation can define `soft modes'  where pinning effects are weak and small forces can lead to sizable effect. 
To analyze such  rotational forces we have projected the widely used Landau Lifshitz Gilbert equation onto the rotational degree of freedom using a straightforward generalization of the approach used by Thiele \cite{Thiele72} for translational motion. We expect that this approach should also be useful to analyze other sources of rotational torques both in skyrmion lattices and other experiments where rotation plays a role.\cite{Khvalkovskiy09}

After this work was completed, a new type of damping term was suggested by Zang {\it et al.} in Ref.~[\onlinecite{zang11}] which complements the Gilbert damping in the LLG equation. This additional damping will modify prefactors in our results for the rotation angle but we do not expect any qualitative changes.

\acknowledgments
We thank especially Christian Pfleiderer and his group for numerous helpful 
discussions. Furthermore we acknowledge
discussion with N. Nagaosa. We thank the DFG for financial support within SFB 608.
K. E. was supported by Deutsche Telekom Stiftung and the Bonn-Cologne graduate
school (BCGS).
R.D. was supported by FOM, NWO and the ERC.
\\

\appendix

\section{Alternative derivation of the rotation angle $\delta \Phi$}
\label{secVar}

The derivation of Eq.~(\ref{delPhi}) for the rotation angle was based on the
projection of the LLG equation on the rotational mode around an axis defined by
the normal vector $\hat{\vect n}$ of the skyrmion lattice, see ~\ref{normal}.
In order to check this approach and confirm the validity of Eq.~(\ref{delPhi}),
we used an alternative derivation {\em without} projection to the rotational
mode. 

In this alternative approach, we have solved for a steady-state solution of the
LLG equation \eqref{LLG} directly within the variational ansatz of
Eq.~(\ref{magnetvar}) with variational parameters $\mu_i$, $i = 1,..,N$ with
$N=27$. In order to determine the change $\delta \mu_i$ of the variational
parameters in the presence of a current $\vect v_s$, we multiply
Eq.~(\ref{LLG1}) by $\partial 
\hat{\vect \Omega} / \partial \mu_i$ and integrate over space. As the effective
magnetic field $\vect H_{\rm eff}$ vanishes in equilibrium, we expand it to
linear order in the deviations $\delta \mu_i$ to obtain $N$ equations 
generalizing Eq.~(\ref{delPhi})
\begin{equation}
\label{TorqueG}
{\vect P}_{R,i} (\vect{v}_s-\vect{v}_d)+
{\vect P}_{D,i} (\beta \vect{v}_s - \alpha \vect v_d)=f''_{ij} \delta \mu_j
\end{equation}
where the generalized reactive and dissipative coupling vectors, ${\vect
P}_{R/D,i}$ and the stiffness matrix $f''_{ij}$ are given by
\begin{subequations}
\begin{align}
({\vect P}_{R, i})_n &=\int_{\rm UC} d^2 r\, 
\left( \hat{\vect \Omega} \times \partial_n \hat{\vect \Omega} \right)
\frac{\partial \hat{\vect \Omega}}{\partial \mu_i}, \label{prf} 
\\ 
(\vect P_{D, i})_n &=\int_{\rm UC} d^2 r\, \partial_n
\hat{\vect \Omega} 
\,\frac{\partial \hat{\vect \Omega}}{\partial \mu_i} \label{pdf}
\\
f''_{ij}&=\int_{\rm UC} d^2 r \, \frac{\partial \hat{\vect \Omega}}
{\partial \mu_i} \frac{ \partial \vect H_{\rm eff}}{\partial \mu_j}=-\frac{1}{M} \int_{\rm UC}  d^2
 r\,
\frac{\partial^2 F}{\partial \mu_i \partial \mu_j} \label{fpp}
\end{align}
\end{subequations}
As discussed in section \ref{RotationalMode}, terms in the integrand that are
linear in the coordinate $\vect r$ arise from derivatives 
with respect to the reciprocal lattice vectors $\vect Q_j$. We again neglect
such terms which is justified for 
symmetric boundary conditions.

The rotation angle $\delta \Phi$ can be obtained from the changes in the 
magnetic texture parameterized by the deviations $\delta \mu_i$ of the 
variational parameters. The rotation is obtained from the change 
$\delta \vect Q_i$ of the reciprocal lattice vectors
of the skyrmion lattice. For small $\delta \Phi$ we have
$\delta \vect Q_i = (\partial \vect Q_i / \partial \Phi)\, \delta \Phi$.
Multiplying this formula by $(\partial \vect Q_i / \partial \Phi)$ and summing 
over $i$ allows to solve for $\delta \Phi$.
Thus, we get the alternative expression for the rotation angle 
\begin{equation} 
\label{deltaPhi}
\delta \Phi =
\frac{\sum_i  \delta \vect{Q}_i \cdot \frac{\partial \vect{Q}_i}{\partial \Phi}}
{\sum_i  \frac{\partial \vect{Q}_i}{\partial \Phi}\cdot \frac{\partial
\vect{Q}_i}{\partial \Phi}}
\end{equation}
The deviations $\delta \vect Q_i$ are obtained by solving Eq.~(\ref{TorqueG})
using Eq.~(\ref{vd}) for the drift velocity $\vect v_d$.

We have checked both analytically and numerically that Eq.~(\ref{deltaPhi})
and Eq.~(\ref{delPhi}) give identical results, and thus confirmed the validity of the Thiele approach in the present
context for the rotational motion.

\bibliography{spintorque}
\bibliographystyle{apsrev}

\end{document}